\documentclass[12pt,nohyper]{JHEP3}
\usepackage{graphics,psfrag}
\usepackage{amsthm,amssymb,epsfig,amsmath,euscript,array,cite}
\usepackage{amsfonts}

\newcounter{multieqs}

\newcommand{\be}{\begin{equation}}
\newcommand{\ee}{\end{equation}}
\newcommand{\eq}[1]{(\ref{#1})}
\newcommand{\bit}{\begin{itemize}}  \newcommand{\eit}{\end{itemize}}

\newcommand{\bm}[1]{\mbox{\boldmath $#1$}}
\newcommand{\rf}[1]{(\ref{#1})}

\def\bd{\begin{document}}
\def\ed{\end{document}}
\def\nn{\nonumber}
\def\bea{\begin{eqnarray}}
\def\eea{\end{eqnarray}}
\let\bm=\bibitem

\def\la{\langle}
\def\ra{\rangle}

\def\npb#1#2#3{Nucl. Phys. {\bf{B#1}} #3 (#2)}
\def\plb#1#2#3{Phys. Lett. {\bf{#1B}} #3 (#2)}
\def\prl#1#2#3{Phys. Rev. Lett. {\bf{#1}} #3 (#2)}
\def\prd#1#2#3{Phys. Rev. {D \bf{#1}} #3 (#2)}
\def\cmp#1#2#3{Comm. Math. Phys. {\bf{#1}} #3 (#2)}
\def\cqg#1#2#3{Class. Quantum Grav. {\bf{#1}} #3 (#2)}
\def\nppsa#1#2#3{Nucl. Phys. B (Proc. Suppl.) {\bf{#1A}}#3 (#2)}
\def\ap#1#2#3{Ann. of Phys. {\bf{#1}} #3 (#2)}
\def\ijmp#1#2#3{Int. J. Mod. Phys. {\bf{A#1}} #3 (#2)}
\def\rmp#1#2#3{Rev. Mod. Phys. {\bf{#1}} #3 (#2)}
\def\mpla#1#2#3{Mod. Phys. Lett. {\bf A#1} #3 (#2)}
\def\jhep#1#2#3{J. High Energy Phys. {\bf #1} #3 (#2)}
\def\atmp#1#2#3{Adv. Theor. Math. Phys. {\bf #1} #3 (#2)}

\def\N{{\cal N}}
\def\sst{\scriptscriptstyle}
\def\thetabar{\bar\theta}
\def\Tr{{\rm Tr}}
\def\one{\mbox{1 \kern-.59em {\rm l}}}

\def\a{\alpha}      \def\da{{\dot\alpha}}  \def\dA{{\dot A}}
\def\b{\beta}       \def\db{{\dot\beta}}  
\def\g{\gamma}  \def\G{\Gamma}  \def\dc{{\dot\gamma}}  
\def\d{\delta}  \def\D{\Delta}  \def\ddt{\dot\delta}  
\def\e{\epsilon}        \def\ve{\varepsilon}  
\def\f{\phi}    \def\F{\Phi}    \def\vvf{\f}  
\def\h{\eta}  
\def\k{\kappa}  
\def\l{\lambda} \def\L{\Lambda}  
\def\m{\mu} \def\n{\nu}  
\def\o{\omega}  
\def\p{\pi} \def\P{\Pi}  
\def\r{\rho}  
\def\s{\sigma}  \def\S{\Sigma}  
\def\t{\tau}  
\def\th{\theta} \def\Th{\Theta} \def\vth{\vartheta}  
\def\X{\Xeta}  
\def\z{\zeta}  

\def\na{\nabla}  

\def\cA{{\cal A}} \def\cB{{\cal B}} \def\cC{{\cal C}}  
\def\cD{{\cal D}} \def\cE{{\cal E}} \def\cF{{\cal F}}  
\def\cG{{\cal G}} \def\cH{{\cal H}} \def\cI{{\cal I}}  
\def\cJ{{\cal J}} \def\cK{{\cal K}} \def\cL{{\cal L}}  
\def\cM{{\cal M}} \def\cN{{\cal N}} \def\cO{{\cal O}}  
\def\cP{{\cal P}} \def\cQ{{\cal Q}} \def\cR{{\cal R}}  
\def\cS{{\cal S}} \def\cT{{\cal T}} \def\cU{{\cal U}}  
\def\cV{{\cal V}} \def\cW{{\cal W}} \def\cX{{\cal X}}  
\def\cY{{\cal Y}} \def\cZ{{\cal Z}}

\def\ua{\underline{\alpha}}  
\def\uc{\underline{\phantom{\alpha}}\!\!\!\gamma}  
\def\um{\underline{\mu}}  
\def\ud{\underline\delta}  
\def\ue{\underline\epsilon}  
\def\una{\underline a}\def\unA{\underline A}  
\def\unb{\underline b}\def\unB{\underline B}  
\def\unc{\underline c}\def\unC{\underline C}  
\def\und{\underline d}\def\unD{\underline D}  
\def\une{\underline e}\def\unE{\underline E}  
\def\unf{\underline{\phantom{e}}\!\!\!\! f}\def\unF{\underline F}  
\def\unm{\underline m}\def\unM{\underline M}  
\def\unn{\underline n}\def\unN{\underline N}  
\def\unp{\underline{\phantom{a}}\!\!\! p}\def\unP{\underline P}  
\def\unq{\underline{\phantom{a}}\!\!\! q}  
\def\unQ{\underline{\phantom{A}}\!\!\!\! Q}  
\def\unH{\underline{H}}  
  
\def\As {{A \hspace{-6.4pt} \slash}\;}  
\def\bs {{b \hspace{-6.4pt} \slash}\;}  
\def\Ds {{D \hspace{-6.4pt} \slash}\;}  
\def\ds {{\del \hspace{-6.4pt} \slash}\;}  
\def\ss {{\s \hspace{-6.4pt} \slash}\;}  
\def\ks {{ k \hspace{-6.4pt} \slash}\;}  
\def\ps {{p \hspace{-6.4pt} \slash}\;}   
\def\xs {{x \hspace{-6.4pt} \slash}\;}  
\def\pas {{{p_1} \hspace{-6.4pt} \slash}\;}  
\def\pbs {{{p_2} \hspace{-6.4pt} \slash}\;}   
\def\cFs {{{\cal F} \hspace{-6.4pt} \slash}\;}

\def\Dh{\hat{D}}
\def\Gh{\hat{G}}
\def\Fh{\hat{F}}
\def\Ph{\hat{P}}
\def\Rh{\hat{R}}
\def\Vh{\hat{V}}  
\def\Xh{\hat{X}} 
 
\def\ah{{\hat{a}}}
\def\gh{\hat{g}} 
\def\hh{\hat{h}}
\def\uh{\hat{u}}  
\def\xh{\hat{x}}  
\def\yh{\hat{y}}  
\def\ph{\hat{p}}  
\def\xih{\hat{\xi}}  
\def\chih{\hat{\chi}}

\def\psit{\tilde{\psi}}  
\def\Psit{\tilde{\Psi}}   
\def\Psibt{\tilde{\bar{Psi}}}  

\def\st{\tilde{\sigma}}  

\def\Phit{\tilde{\Phi}}   
\def\Phitb{\overline{\tilde{Phi}}}  
\def\tht{\tilde{\th}}  
\def\lt{\tilde{\l}}
\def\chit{\tilde{\chi}}   
\def\phit{\tilde{\phi}} 

\def\At{\tilde{A}}
\def\Bt{\tilde{B}}
\def\Ct{\tilde{C}}
\def\Dt{\tilde{D}}
\def\Et{\tilde{E}}
\def\Ft{\tilde{F}}
\def\Ht{\tilde{H}}
\def\Qt{\tilde{Q}}  
\def\Rt{\tilde{R}}  
\def\Mt{\tilde{M }}  
\def\Nt{\tilde{N}}   
\def\St{\tilde{S}}
\def\Vt{\tilde{V}}
\def\Xt{\tilde{X}} 
\def\at{\tilde{a}}
\def\ct{\tilde{c}}   
\def\htt{\tilde{h}} 
\def\ft{\tilde{f}}
\def\gt{\tilde{g}}
\def\pt{\tilde{p}}  
\def\qt{\tilde{q}}  
\def\vt{\tilde{v}}  
\def\nt{\tilde{n}}  
\def\ut{\tilde{u}}  
\def\wt{\tilde{w}}  
\def\zt{\tilde{z}} 
\def\xt{\tilde{x}} 
\def\yt{\tilde{y}} 
\def\Psit{\tilde{\Psi}}
\def\vphit{\tilde{\varphi}}  

\def\delb{\bar{\partial}}  
\def\thb{\bar{\theta}}
\def\mub{\bar{\mu}}
\def\lamb{\bar{\l}}
\def\psib{\bar{\psi}}
\def\sb{\bar{\sigma}}
\def\xib{\bar{\xi}}
\def\chib{\bar{\chi}}

\def\Phib{\bar{\Phi}}
\def\Lamb{\bar{\Lambda}}
\def\Sb{{\overline \Sigma}}
\def\cb{\bar{c}}
\def\wb{\bar{w}}
\def\ub{\bar{u}}
\def\zb{{\bar{z}}}
\def\Qb{{\bar Q}}
\def\qb{\bar{q}}

\def\Ab{{\overline A}} \def\Bb{{\overline B}} \def\Cb{{\overline C}}  
\def\Db{{\overline D}} \def\Eb{{\overline E}} \def\Fb{{\overline F}}  
\def\Gb{{\overline G}} \def\Hb{{\overline H}} \def\Ib{{\overline I}}  
\def\Jb{{\overline J}} \def\Kb{{\overline K}} \def\Lb{{\overline L}}  
\def\Mb{{\overline M}} \def\Nb{{\overline N}} \def\Ob{{\overline O}}  
\def\Pb{{\overline P}}  \def\Rb{{\overline R}}  
 \def\Tb{{\overline T}} \def\Ub{{\overline U}}  
\def\Vb{{\overline V}} \def\Wb{{\overline W}} \def\Xb{{\overline X}}  
\def\Yb{{\overline Y}} \def\Zb{{\overline Z}}  

\def\fb{{\overline f}}
\def\gb{{\overline g}}
\def\mb{{\overline m}}
\def\lb{{\overline l}}
\def\yb{{\overline y}}

\def\ba{{\bf a}} 
\def\bk{{\bf k}}  
\def\bl{{\bf l}}  
\def\bp{{\bf p}}  
\def\bq{{\bf q}}  
\def\br{{\bf r}}
\def\bt{{\bf t}}
\def\bu{{\bf u}}
\def\bv{{\bf v}}
\def\bx{{\bf x}}  
\def\by{{\bf y}}  
\def\bR{{\bf R}}  
\def\bV{{\bf V}}

\def\bone{{\bf 1}}  

\def\va{{\vec a}}
\def\vk{{\vec k}}
\def\vp{{\vec p}}
\def\vq{{\vec q}}
\def\vx{{\vec x}}
\def\vy{{\vec y}}
\def\vu{{\vec u}}
\def\vv{{\vec v}}

\def\vs{{\vec \sigma}}
\def\vtau{{\vec \tau}}

\newcommand{\ov}[1]{\overrightarrow{#1}}

\def\mA{\mathfrak{A}}
\def\mB{\mathfrak{B}}
\def\mC{\mathfrak{C}}
\def\mD{\mathfrak{D}}
\def\mE{\mathfrak{E}}
\def\mF{\mathfrak{F}}
\def\mG{\mathfrak{G}}
\def\mH{\mathfrak{H}}
\def\mM{\mathfrak{M}}
\def\mN{\mathfrak{N}}
\def\mW{\mathfrak{W}}

\def\ma{\mathfrak{a}}
\def\mb{\mathfrak{b}}
\def\mf{\mathfrak{f}}
\def\mg{\mathfrak{g}}
\def\mh{\mathfrak{h}}
\def\ml{\mathfrak{l}}

\def\d{\delta}\def\D{\Delta}\def\ddt{\dot\delta}  
  
\def\pa{\partial} \def\del{\partial}  
\def\xx{\times}  
\def\uno{\mbox{1 \kern-.59em {\rm l}}}    
  
\def\trp{^{\top}}  
\def\inv{^{-1}}  
\def\dag{{^{\dagger}}}  
\def\pr{^{\prime}}  
  
\def\rar{\rightarrow}  
\def\lar{\leftarrow}  
\def\lrar{\leftrightarrow}  
  
\newcommand{\0}{\,\!}      
\def\one{1\!\!1\,\,}  
\def\im{\imath}  
\def\jm{\jmath}  
  
\newcommand{\tr}{\mbox{tr}}  
\newcommand{\slsh}[1]{/ \!\!\!\! #1}  
  
\def\vac{|0\rangle}  
\def\lvac{\langle 0|}  
  
\def\hlf{\frac{1}{2}}  
\def\ove#1{\frac{1}{#1}}  

\def\Box{\square}  
\def\CC {\mathbb{C}}
\def\RR{\mathbb{R}}
\def\ZZ{\mathbb{Z}}  
\def\bb#1{{\bf #1}}  
\def\bcomment#1{}  
\def\bfhat#1{{\bf \hat{#1}}}  
\def\VEV#1{\left\langle #1\right\rangle}  

\newcommand{\ex}[1]{{\rm e}^{#1}} \def\ii{{\rm i}}  

\newcommand{\lrbrk}[1]{\left(#1\right)}
\newcommand{\sfrac}[2]{{\textstyle\frac{#1}{#2}}}
 
\def\stw{{\sqrt{2}}}

\def\rf {{\rm f}}
\def\ri {{\rm i}}
\def\rs {{\scriptscriptstyle \rm S}}
\def\rt {{\scriptscriptstyle \rm T}}

\def\rQ {{\scriptscriptstyle \rm \cQ}}
\def\rR {{\scriptscriptstyle \rm \cR}}

\def\cQb{{\cal \Qb}}
\def\cRb{{\cal \Rb}}
\def\cWb{{\cal \Wb}}

\def\fd {{\rm N}}
\def\afd {{\overline{\rm N}}}

\def \II {I\hspace{-.1em}I\hspace{.1em}}
\def \IIA {\mbox{\II A\hspace{.2em}}}
\def \IIB {\mbox{\II B\hspace{.2em}}}
\def \gs {g^s}
\def \ls {\lambda^s}

\def \I {{\cal I}}
\def \qs {q\hspace{-.53em}/\hspace{.15em}}
\def \ks {k\hspace{-.53em}/\hspace{.15em}}
\def \YM {{\mbox{\tiny YM}}}
\def \gym {g_{\YM}}

\def \Lc {\L_c}
\def\IR{\relax{\rm I\kern-.18em R}}
\def \id {{\bf 1}}
\def \Aa { {A^{(1)}} }
\def \Ab { {A^{(2)}} }

\author{Chong-Sun Chu and Douglas J.   Smith  \\  
Centre for Particle Theory
and Department of Mathematics, 
Durham University, Durham, DH1 3LE, UK \\
E-mail:  
\email{chong-sun.chu@durham.ac.uk, Douglas.Smith@durham.ac.uk} }

\title {Multiple Self-Dual Strings on M5-Branes}

\abstract{ We show how to define Chern-Simons matter theories with
boundary. Rather than imposing boundary conditions, we introduce new
boundary degrees of freedom from the beginning and show how they can be
used to cancel the gauge non-invariance of the Chern-Simons action. 
We apply this method to the ABJM theory with boundary. By imposing also
boundary conformal invariance, we determine the required boundary
action. This result allows us to derive the action for the multiple
self-dual strings living on  M5-branes.
}
 
\preprint{DCPT-09/63}
\keywords{Chern-Simons Theories, M-Theory, Anomalies in Field 
and String Theories, M(atrix) Theories}

\begin{document}

\section{Introduction}

M-theory contains M2-branes and M5-branes. The understanding of the
dynamics and interactions of these branes are some of the most important
and mysterious aspects of M-theory. Recent progress has been made in the
description of multiple M2-branes through the work of Bagger and Lambert
\cite{BLG1,BLG2,BLG3} and Gustavsson \cite{BLG4}. The Bagger-Lambert (BL)
theory \cite{BLG1,BLG2,BLG3} was originally motivated by trying to
construct an action with manifest ${\cal N} = 8$ superconformal
symmetry, based on a BPS equation postulated by Basu and Harvey
\cite{BH}. This naturally led to an action with a non-abelian symmetry
based on a Lie 3-algebra. However there is only one example of such a
3-algebra and it has been difficult to increase the rank of the gauge group.

Aharony, Bergman, Jafferis and Maldacena (ABJM) \cite{ABJM} 
proposed a $U(N)\times
U(N)$ Chern-Simons gauge theory with levels $k$ and $-k$  arises on the
worldvolume of $N$ M2-branes placed at the orbifold singularity
$\RR^8/\ZZ_k$. The theory allows arbitrary rank, but 
has only $\cN =6$ supersymmetry, although it is
conjectured that supersymmetry is enhanced to $\cN=8$ for $k=1,2$. See
for example \cite{rey} for some recent discussions on this issue.

The worldvolume theory on a flat M5-brane is given by a six dimensional (2,0) 
superconformal field theory. The massless excitations are given by the tensor
multiplet which consists of a self-dual 2-form potential, five scalar
fields and 8 fermions. These fields are related to the breaking of the
symmetries of the 11 dimensional supergravity, namely the 
gauge symmetry of the three-form potential, the 11 dimensional  translational
invariance, and the supersymmetries.
For $(r+1)$ M5-branes parallel to each other, one obtains the $A_r$ series
of (2,0) theory \cite{strominger,witten-orbifold}. 
The existence of these theories were first argued for 
\cite{witten-comments}
by considering type IIB string theory compactified on a $K_3$ with a
2-cycle shrinking to zero size and developing an
$A_r$ type singularity. The resulting IIB string
theory is tensionless and is intrinsically non-perturbative, making
it hard to study. These
strings can also be viewed as the boundaries of open M2-branes which
end on M5-branes. As tensionless self-dual strings 
are difficult to understand, one approach to the problem is to introduce
a vev for one of the scalars which results in self-dual strings with
finite tension. In the M-theory picture, turning on the vev 
corresponds to separating the M5-branes. A comprehensive review of this
approach can be found in the thesis \cite{arv}.

Just as strings can end on D-branes,  M2-branes can end on M5-branes.
It is natural to try to use the open M2-branes system to learn 
about the physics of the M5-brane or about the intersection itself,
i.e. the self-dual string theory. 
Recently, by considering  a system of
M2-branes ending on an M5-brane with a constant $C$-field turned on, 
the quantum geometry on the M5-brane worldvolume has been derived
\cite{CS1}. In this paper, we will consider a system of open M2-branes
and use it to learn about the physics of multiple self-dual strings.

For a single self-dual string, the action is given by a free part,
plus a coupling to the self-dual tensor potential field, all
constructed in a superconformal fashion \cite{arv}. In this paper, we
are
interested in constructing the bosonic part of the 
theory for multiple self-dual strings using the ABJM theory.
In analogy to the usual matrix string theory \cite{ms},
one expects to promote the string coordinates to
matrices and introduce a self-interaction among them.  What is new
in the ABJM
theory is the presence of a $U(N)\times U(N)$ twisted Chern-Simons
action. 
Twisted here means the levels of the two gauge groups of the Chen-Simons
action are opposite in sign.
The action is not gauge invariant when there is a
boundary. This leads to interesting new consequences for the dynamics
of the self-dual strings theory.
 
The canonical way to deal with the
gauge non-invariance for a standard Chern-Simons action is by
imposing a boundary condition \cite{witten-jones,emss}. The boundary
condition breaks gauge invariance at the boundary, and hence 
extra degrees of freedom reside there. Since there are two factors of
$U(N)$ gauge group, the new extra degrees of freedom 
take values in each factor of the $U(N)$. 
This is well understood for ordinary Chern-Simons theory. For the ABJM
theory, there are three points one has to be careful about. First, the
Chern-Simons terms come with opposite levels. 
In \cite{Ber1,Ber2},  a straightforward
application of the ``imposing boundary condition'' procedure was
adopted. The authors found
opposite kinetic terms for the new degrees of freedom and so 
the boundary theory considered there is non-unitary. Secondly,
the theory is no longer topological due to the presence of
matter. Therefore even though one can impose boundary conditions on
the gauge fields to make the gauge non-invariance of the action
vanish, it is not clear in this approach how the expected degrees of
freedom would arise (see the discussions in the last paragraph of sec
2.3). Finally, if we consider a boundary ABJM theory which preserves part of the
bulk supersymmetry, the boundary degrees of freedom will have to form
a supermultiplet
of the boundary supersymmetry. This will provide strong constraints on the
boundary degrees of freedom.

In this paper, we demonstrate a new way to treat the gauge
non-invariance created by the
boundary Chern-Simons action. For pure untwisted Chern-Simon theory,
our methods gives identical results to the original treatments of
\cite{witten-jones,emss}. However our method is applicable also for
twisted Chern-Simons theory with matter, e.g. the ABJM theory.
We will see that this  gauge non-invariant term is analogous to 
the gauge anomaly
in many ways, so we will call this an anomaly.
We find that apart from imposing a boundary condition, one can 
do away with the anomaly
by introducing a set of extra boundary degrees of
freedom whose action has a variation which cancels the anomaly. 
Not surprisingly, 
the required action is the Wess-Zumino
term \cite{WZ}. Our construction is
modeled after the original construction of Wess and Zumino
\cite{WZ}. The extra degrees of freedom play the role of the Nambu-Goldstone
bosons. 
In addition to gauge invariance, 
the 
boundary ABJM theory 
is also expected to be conformally
invariant. This can be achieved by adding  additional kinetic terms to
the theory. This results in a Wess-Zumino-Witten 
(WZW) theory. 

The plan of the paper is as follows. In section 2, we first review the
original treatment of \cite{witten-jones,emss} for a pure untwisted
boundary Chern-Simons action. Then we explain our new method in section
2.2. A comparison of the two methods is given in section 2.3, where we
demonstrate that the two are equivalent for the pure Chern-Simon case.
In section 3.1, we apply our construction to the bosonic ABJM theory with a boundary.
We argue for
and identify the extra degrees of freedom that must be present on the
boundary of the open M2-branes system. 
The gauge invariant and conformal invariant bosonic action  is
determined.
In section 3.2 we take into account the boundary supersymmetry and describe how
the bosonic field content has to be enlarged. The resulting supersymmetric
WZW model for these degrees of freedom is presented.
Then, in section 3.3, 
by considering a specific configuration of open membranes 
suspended between two M5-branes, we determine the 
action for a
system of multiple self-dual strings. The paper concludes with
discussions in section 4.

\section{Chern-Simons  Theory and Boundary Action}

Consider a Chern-Simons  theory with gauge group $G$ on a 3-dimensional 
manifold $M$
\be
S_{CS} = \frac{k}{4 \pi} \int_M \omega_3(A) 
= \frac{k}{4 \pi} \int_M \Tr ( A d A + \frac{2}{3}A^3 ) ,
\ee
where $A^3$ denotes $A \wedge A \wedge A$ etc and 
$A = dx^\m A_\m$ is a Lie algebra valued one-form.
When $M$ is closed, the theory is gauge invariant and topological.
In general the Chern-Simons form $\o_{2n+1}(A)$ satisfies 
\be
\Tr F^{n+1} = d \o_{2n+1} (A), \quad \mbox{where $F = dA +
A^2$}. 
\ee
Under an infinitesimal gauge transformation
\be
\d_\a A = d \a + [A, \a], \quad \d_\a F = [F,\a], 
\ee
the Chern-Simons form transforms as
\be \label{des}
\d_\a \o_{2n+1}(A) = d \o_{2n}^1 (A; \a).
\ee
Explicit expressions for $\o_{2n+1}(A)$ and  
$\o_{2n}^1 (A; \a)$ can be
computed using the Cartan homotopy operator, see \cite{z1} for details.
For example, 
\be
\o_2^1 (A; \a) = \Tr(\a dA). 
\ee

To study the theory in the presence of a boundary $\partial M$, we note that
due to \eq{des} 
the action $S_{CS}$ is not gauge invariant, but its variation is a boundary
term.   Hence the gauge invariance of the action $S_{CS}$ is
broken at the boundary.
These boundary terms will vanish with appropriate boundary conditions
\cite{witten-jones,emss}. We will review this approach in the next subsection.
Alternatively they can be cancelled by the variations of additional degrees of
freedom. We will explain this new approach in subsection 2.2.

\subsection{Imposing boundary conditions}

Consider an arbitrary infinitesimal variation of $A$, the variation of the
Chern-Simons action  gives
\be \label{bdy-var}
\d S_{CS} = \frac{k}{2\pi}\int_M \Tr( \d A F) + 
\frac{k}{4\pi}\int_{\partial M} \Tr \left( \delta A A\right) .
\ee
The bulk term gives the equation of motion $F=0$. 
For the boundary term to vanish, one can impose a boundary condition on $A$
and only allow variations which preserve the boundary condition. The most
general condition is a linear relation between the two boundary
components of $A$. Each choice of the boundary condition
corresponds to a definition of the boundary theory. In general,  one can 
divide the possible boundary conditions into different inequivalent classes, 
each
corresponding to a possible definition of the boundary theory. 

For physical applications, we consider manifolds of the
form $M= \RR \times \Sigma$, where the noncompact direction $\RR$ is 
interpreted as
time and $\del \S \neq 0$ \cite{emss}. 
Let's consider explicitly a boundary at $x^2 = 0$ and
choose the boundary condition $A_0 = 0$.
With this boundary condition,  one can write
\be \label{S2}
S_{CS} = \frac{k}{2\pi} \int_M \Tr (\e^{ij}F_{ij} A_0 
- \frac{1}{2} \e^{ij} A_i \dot{A}_j),
\ee
where we have integrated by parts the terms involving a derivative of
$A_0$, whose resulting boundary terms vanishing either due to the
usual asymptotic boundary conditions, or because $A_0$ vanishes on $\partial M$.
The result is that $A_0$ only appears linearly in \eq{S2}, and is therefore
a Lagrange multiplier, imposing the constraint $F_{12} = 0$. This means that
we can write
\be \label{AU}
A_i = U^{-1}\partial_i U \;\; {\rm for} \;\; i = 1,2 
\ee
for $U \in G$. 
If we then substitute this back into the action \eq{S2} we find
\be
S = - \frac{k}{8\pi}\int_{\partial M} \Tr 
\left( U^{-1} \del_0 U U^{-1} \del_1 U \right)
- \frac{k}{12\pi}\int_M \Tr \left( U^{-1}dU \right)^3  .
\ee
Note that a ``chiral kinetic term'' is obtained for $U$.
A few remarks follow.

\bit
\item[1.]  Note  that this result is essentially the same for any
allowed boundary conditions for $A$. However, the form of the 
boundary (kinetic) term
does depend on whether we choose a timelike, spacelike or lightlike combination
of components of $A$ to vanish on the boundary. For example, if we had
taken the boundary condition  $A_1 = 0$ instead 
of $A_0 = 0$, then the sign of the kinetic term would be reversed. 
\if 
Clearly the
relative sign between the bulk and boundary terms depends on the orientation
of the boundary. However, for a fixed orientation, the sign of the kinetic term 
would change if
we chose a spacelike rather than a timelike combination of components of $A$
to vanish, 
\fi
Furthermore, if
we instead chose a light-like combination $A_0 \pm A_1 = 0$ on the boundary,
we would get a ``conventional kinetic term'' 
$( U^{-1} \partial_{\mu} U )^2$ on the boundary, 
with the overall sign depending on which light-like
direction we chose.

So, choosing appropriate boundary conditions for $A$, we arrive at the
well-known WZW action 
\be \label{wzw1}
S_{WZW}[U] = 
  - \frac{k}{8\pi}\int_{\partial M} \Tr ( U^{-1}\partial_{\mu}U)^2 
- \frac{k}{12\pi}\int_M \Tr \left( U^{-1}dU \right)^3,
\ee
where the metric is $\eta_{00} =-1 = - \eta_{11}$.
This action describes the dynamics for the field $U$
living  on the boundary $\del M$. 
 
\item[2.] 
We remark that 
while the above derivation is classical, it was argued in \cite{emss} that
there is no non-trivial Jacobian introduced in the path integral by the
change of variables \eq{AU}. 
Therefore the action in terms of $U$ is equivalent to
the original Chern-Simons action with boundary conditions on $A$.

\item[3.] 
We note that although the boundary conditions 
break the original gauge symmetry,
the resulting WZW action \eq{wzw1} 
does have a symmetry \cite{W1}. In fact the action is invariant
under chiral transformations of the form
\be
U \rightarrow \Omega(z) U \tilde{\Omega}(\bar{z})
\ee
where $z = x^0 + ix^1$ and $\Omega, \tilde{\Omega} \in G$. This 
chiral $G\times G$
symmetry is generated by the currents ($\del= \del_z$, $\delb =
\del_{\zb}$)
\be
J = \frac{k}{\pi} U^{-1} \del U, \quad 
\bar{J} = \frac{k}{\pi} \delb U U^{-1}, 
\ee
which are
chiral, $\del \bar{J} = \delb J =0$ and satisfy a
Kac-Moody algebra
\bea
\; [J^a (z_1), J^b (z_2)] = i f^{ab}_c J^c(z_1) \d(z_1-z_2) 
- i \frac{k}{\pi} \d'(z_1-z_2) \d^{ab}, \nn\\
\; [\bar{J}^a (\zb_1), \bar{J}^b(\zb_2)] = i f^{ab}_c \bar{J}^c(\zb_1) 
\d(\zb_1-\zb_2) 
+ i \frac{k}{\pi} \d'(\zb_1-\zb_2) \d^{ab}.
\eea
We remark that the Kac-Moody symmetry 
is not part of the original gauge symmetry of the
Chern-Simons action.

\item[4.] 
We also note that the action \eq{wzw1}
is conformal -- for this to hold at the quantum level the relative
coefficient between the bulk and boundary terms is fixed, as above, up to a
sign \cite{W2}. The fact that a conformal field theory arises on the
boundary can be understood as follows: we start out with the bulk
Chern-Simons theory which is topological, i.e. invariant under 
arbitrary variations 
of the metric. 
By imposing a boundary condition, a conformal structure is fixed 
on the boundary manifold. Therefore the boundary theory can only be 
invariant under arbitrary variations of the metric which preserve 
this conformal structure. Hence the boundary theory is conformally invariant. 
\eit

\subsection{Boundary degrees of freedom}
\label{CSBDoF}

Another point of view is that to
render the theory gauge invariant, there should arise additional physical
degrees of freedom at the boundary. The total action should be chosen so that
the gauge variation of the additional terms cancels the gauge
variation of the original action $S_{CS}$. 

To do this, let us note it is well known that the object $\o^1_{2n} (A;
\a)$ is related to the chiral anomaly in a $2n$-dimensional gauge
theory. One of the consequences of the anomaly is that the low energy
effective theory of the Nambu-Goldstone bosons admits a Wess-Zumino term
\cite{WZ}. For the simplest case where the whole gauge group $G$ is
broken by an anomaly, the Nambu-Goldstone bosons $g = e^{- \xi}$ are $G$-valued 
and the Wess-Zumino effective action is given by \cite{WZ}, 
\be \label{WZ-eff}
W[\xi, A] := \int_0^1 dt \;   G_\xi [A_t],
\ee
where $A_t$ is the gauge field $A_t := e^{t \xi} A e^{-t \xi} + e^{t
\xi} d e^{- t \xi}$ and $G_\a[A]= \int \o_{2n}^1(A;\a)$ is the anomaly.  
The effective action was constructed to
reproduce the anomaly: $\d_\a W[\xi,A] =  G_\a [A]$.
It is not difficult to show that \cite{CHZ}
\footnote{ The paper \cite{CHZ} deals with the general situation 
of a gauge theory with gauge group $G$ and
with a subgroup $H$ such that the associated currents are anomaly free.
In this case the Wess-Zumino term can be constructed from the Nambu-Goldstone
bosons which are valued in the coset space $G/H$.
To apply to the present case with trivial $H$, we just need to set $A_h =0$ in
the formula in eqn (115) of appendix 2 there. 
}
\be \label{WZ-eff2}
W[\xi, A] = \int_{M} (- \o_{2n+1}(A^g) + \o_{2n+1}(A) ),
\ee
where $M$ is a $(2n+1)$-dimensional manifold whose boundary $\del M$ is
equal to the $2n$-dimensional spacetime
and $A^g$ is the transform of $A$ under a finite gauge
transformation
\be
A^g := g^{-1} A g + g^{-1} d g.
\ee
Note that since
$d\o_{2n+1}(A) = \tr F^{2n+2}(A)$, the integrand is closed
and so 
\eq{WZ-eff2}  actually defines a $2n$-dimensional action on $\del M$. 
If one expands \eq{WZ-eff2}
around $A=0$, then one finds the WZW term  \cite{W2}
\be
\int_{M} \Tr\; (g^{-1} d g)^{2n+1}.
\ee

Back to our case of a 3-dimensional 
Chern-Simons theory with a boundary. The gauge
non-invariance of the Chern-Simons action is given by 
$\frac{k}{4\pi} \int_{\del M}
\o_2^1(A;\a)$ and is precisely of the same form as the chiral anomaly
reviewed above. Therefore in order to restore gauge invariance, one can
introduce additional degrees of freedom $g$ that live on the boundary (this
plays the role of the Nambu-Goldstone bosons) with the following action
\be
S_{Bdry}:= \frac{k}{4 \pi} \int_M \left[ \omega_3(A^g) - \omega_3(A)
\right],
\ee
so as to cancel against the gauge noninvariant terms resulting from the
Chern-Simons action.
Since $S_{Bdry}$ defines an action on
the boundary manifold $\del M$, one can add it to  the Chen-Simons term. 
Under a gauge transformation with parameter $h$, 
\be
A^g \rightarrow (A^g)^h = A^{hg}.
\ee
Therefore  $A^g$ and hence $\o_3(A^g)$ will be gauge invariant
under the combined transformation
\be
A \rightarrow A^h \;\; , \;\; g \rightarrow h^{-1}g.
\ee
The term $S_{Bdry}$ thus has a gauge variation which cancels exactly
that of $S_{CS}$ and so the total action 
\be
S_T := S_{CS} + S_{Bdry}
\ee
is gauge invariant.

Note that the resulting action is not unique. We are free to add any further
gauge-invariant boundary terms. A natural choice is to try to preserve as
much as possible of the original symmetries of the bulk action. Since the
original action was topological, we can at least try to preserve conformal
invariance on the boundary.
Now, if we set the gauge field $A=0$ we are left with
\be
S_{Bdry}[A=0] = -\frac{k}{12\pi} \int_M \Tr \left( g^{-1}dg \right)^3 .
\ee
We can introduce a boundary term giving a kinetic term for the boundary field
$g$, resulting in the well-known WZW conformal field theory
\be
S_{WZW}[g] = -\frac{k}{8\pi} \int_{\partial M} 
\Tr ( g^{-1}\partial_{\mu}g )^2 
-\frac{k}{12\pi} \int_M \Tr \left( g^{-1}dg \right)^3 .
\ee
Restoring the gauge field $A$ we can maintain both gauge 
and conformal invariance by adding boundary terms as above, 
provided we replace the partial derivative $\partial_{\mu}$ with 
a covariant derivative $D_{\mu} = \partial_{\mu} + A_{\mu}$. 
We therefore have the total
boundary action
\bea \label{SBdry}
S_{Bdry} & = & -\frac{k}{8\pi} \int_{\partial M} 
\Tr (g^{-1}D_{\mu}g )^2 + \frac{k}{4\pi} 
\int_M [\omega_3(A^g) - \o_3(A)]\nn \\
&=& S_{WZW}[g] + \frac{k}{4 \pi} \int_{\del M} \del_+ g g^{-1} A_- 
- \frac{k}{8 \pi} \int_{\del M} A_\m^2, 
\eea
where $\del_\pm := \del_0 \pm \del_1$ and the boundary metric is 
$\eta_{00} = -1 = -\eta_{11}$.
The 2-dimensional action \eq{SBdry} describes the dynamics of 
the group-valued degrees
of freedom $g$ in interaction with an external gauge field.
We remark that it is not the same as the standard gauged
WZW   action  
\be
S_{{\rm gauged}\; WZW} := S_{WZW} + \frac{k}{4 \pi} \int_{\del M} 
(A_+ \del_- g g^{-1} -A_- g^{-1} \del_+ g + A_+ g A_- g^{-1} -A_- A_+ ). 
\ee
In particular, our $S_{Bdry}$ is invariant under the gauge
transformation $A_\m \to h^{-1} A_\m h + h^{-1} \del_\m h$, 
$g \to h^{-1} g$; while $S_{{\rm gauged}\; WZW}$ is invariant
under a different gauge transformation for $g$: $g \to h^{-1} g h$. 

Adding the boundary action $S_{Bdry}$ to $S_{CS}$, the total action is 
\be \label{S1}
S_T =  -\frac{k}{8\pi} \int_{\partial M} 
\Tr ( g^{-1}D_{\mu}g )^2 + \frac{k}{4\pi} \int_M \omega_3(A^g).
\ee

Note that above we introduced the kinetic term for $g$ with a specific
coefficient so that the action in conformal at the quantum level, not just
classically. However, this requirement does not fix the sign of the kinetic
term and so for any given sign of $k$, we are free to choose the kinetic term
to have the correct sign. This observation will be important when describing the
boundary extension of the ABJM action where there are two Chern-Simons terms
with opposite levels.

Let us now derive the
boundary conditions of this action, 
which are imposed on-shell as boundary equations of motion.
This can be derived by concentrating on the boundary contributions to $\d S$.
We note that some care must be taken. Since $A^g$ contains derivatives of
$g$, the variation $\delta A^g$ in the bulk, when expressed in terms of
$\delta A$ and $\delta g$ will also give a boundary
contribution involving $\delta g$.

Consider first varying $A$. Since $g^{-1} D_\m g= A_\m^g$, 
the variation of the boundary term in \eq{S1}
is
\be
-\frac{k}{4\pi} \int_{\partial M} \Tr \left( A^{g\;\mu} \delta A^g_{\mu}
\right);
\ee
while, noting that $\delta A^g = g\delta A g^{-1}$ does not contain
derivatives of $\delta A$, the boundary
contribution in the variation of the bulk term in \eq{S1} is contained in
\be
\frac{k}{4\pi} \int_{M} \Tr \left[ A^g d \left( \delta A^g \right) \right] \sim
-\frac{k}{4\pi} \int_{\partial M} \Tr \left( A^g\delta A^g \right) .
\ee
Combining these two contributions we find the boundary term
\be
-\frac{k}{4\pi} \int_{\partial M} \Tr \left( A^g_{+} \delta A^g_{-} \right) .
\ee
Therefore we obtain from $\d A$ the boundary condition
\be
A^g_{+} = 0
\ee
which is equivalent to
\be \label{A+}
A_{+} = - (\partial_{+}g) g^{-1}.
\ee

Next we consider the variation of $g$, noting that now
\be
\delta A^g = - g^{-1} \delta g g^{-1} (A g + dg) 
+ g^{-1} A \delta g + g^{-1} d(\delta g)
\ee
contains derivatives of $\delta g$.
The variation of the boundary term is
\be \label{vSb1}
-\frac{k}{4\pi} \int_{\partial M} \Tr \left( A^{g\;\mu} \delta A^g_{\mu}
\right),
\ee
while the variation of the bulk term gives the following 
contribution to the boundary variation
\bea
 & & \frac{k}{4\pi} \int_{M} \Tr \left( \delta A^g d A^g 
+ A^g d \left( \delta A^g \right) + 2 A^g A^g \delta A^g \right) \nonumber \\
 & = & -\frac{k}{4\pi} \int_{\partial M} \Tr \left( A^g \delta A^g \right) +
	\frac{k}{2\pi} \int_{M} 
\Tr \left[ \left( dA^g + A^g A^g \right) \delta A^g \right] .
\eea
The explicit boundary terms cancel that of \eq{vSb1} since $A^g_{+} = 0$;
while the final bulk term gives a boundary contribution from
$\delta A^g \sim g^{-1} d(\delta g)$,
\be
-\frac{k}{2\pi} \int_{\partial M} \Tr \left( F^g g^{-1} \delta g \right) 
= -\frac{k}{2\pi} \int_{\partial M} \Tr \left( g^{-1} F \delta g \right) .
\ee
Therefore we find  from $\d g$ the boundary condition 
$F_{\m\n}=0$, $\m,\n =0,1$. 
Together with \eq{A+}, we obtain 
\be
A_\m = - (\partial_{\m}g) g^{-1}, \quad \m =0,1.
\ee

\if
Under arbitrary variations $\delta A$ and $\delta g$ we find
\be
\delta S = \frac{k}{2\pi} \int_M \Tr \left[ \delta A \wedge F \right] +
 \frac{k}{4\pi} \int_{\partial M} 
\Tr \left[ \delta A_{-} g A^g_{+} g^{-1} 
- \delta g (\partial_{+} A^g_{-}) g^{-1} \right]
\ee
where clearly as required there are no bulk terms involving the variation of
the boundary field $g$. We see that the (gauge invariant) boundary equations of
motion are $A^g_{+} = 0$ and $\partial_{+} A^g_{-} = 0$, with the
latter condition being equivalent to $F_{-+} = 0$ when $A^g_{+} = 0$.
\fi

Summarizing, the result is that in $M$ we have equations of
motion $F=0$ so that $A$ is pure gauge. The boundary equations of motion, give
the boundary conditions $A = -dg g^{-1}$, and so we have $A = -dg g^{-1}$ in $M$
where $g$ is an arbitrary extension of the boundary field $g$ into $M$.

\subsection{Comparing the two approaches}

To compare with the alternative method of imposing a boundary condition for
$A$, we note that $A_{+}$ appears linearly in $S_T$ and no derivatives of it is 
involved. This can be seen by rewriting the Chern-Simons action in the
following form
\be \label{SCS2}
S_{CS} = \frac{k}{4 \pi} \int_M \Tr \;  A_+ F_{2-} + \frac{k}{8 \pi} \int_M 
\Tr\; (A_2 \del_+ A_- -  A_- \del_+ A_2) 
- \frac{k}{8 \pi} \int _{\del M} \Tr A_- A_+ ,
\ee
where $A_\pm = A_0 \pm A_1$, $dx^{\pm} = (dx^0\pm dx^1)/2$ etc. The last
term in \eq{SCS2}  cancels precisely the last term in \eq{SBdry} and so
\bea
S_T= S_{WZW}[g] &+& \frac{k}{4\pi} \int_{\del M} \Tr\; \del_+ g g^{-1} A_- +
\frac{k}{4 \pi} \int_M \Tr\;  A_+ F_{2-} \nn\\
&+& \frac{k}{8 \pi} \int_M  \Tr\; (A_2 \del_+ A_- -A_- \del_+ A_2).
\eea
As a result,   $A_{+}$ is a (bulk) Lagrange multiplier and
can be integrated out, imposing the constraint $F_{2-} = 0$ in $M$. Solving this
constraint by writing $A_2 = \lambda^{-1} \partial_2 \lambda$ and
$A_{-} = \lambda^{-1} \partial_{-} \lambda$, we find
\be
S_T = S_{WZW}[g] + S_{WZW}[\lambda] + 
\frac{k}{4\pi} \int_{\partial M} 
\Tr \left( \partial_{+}g g^{-1} \lambda^{-1} \partial_{-} \lambda \right) 
= S_{WZW}[\lambda g],
\ee
where we have used the Polyakov-Wiegmann identity in the last step. 

Therefore with the identification $U = \lambda g$, we finally arrive 
at the same action as obtained by the boundary condition approach. 
The degree of
freedom $\l$ which arises from the gauge field $A$ and the boundary
degree of freedom $g$ naturally combine into the variable $U$ which
appears in the boundary condition approach. 
Physically this is expected since the Chern-Simons theory without boundary 
is topological and describes a flat-connection over $M$. This 
corresponds to a single gauge function $U$ when projected to the
boundary. 

Note that by imposing a boundary condition, we break the gauge symmetry,
while by adding boundary degrees of freedom $g$ we preserve the gauge
symmetry.  However  as mentioned already, 
after integrating out the Lagrange multiplier,
$g$ and $\l$ combine into a single variable $U= \l g$. This variable is a gauge
singlet (since gauge transformations transform $g \rightarrow h^{-1}g$ together
with $\lambda \rightarrow \lambda h$) and so the gauge symmetry
is not apparent when we use the description in terms of $U$. 

In the next section we will deal with the boundary ABJM theory with
new features that one needs to be careful with. In addition to a twisted 
$U(N)\times U(N)$ Chern-Simons action, the presence of matter brings
in a new complication. Due to the matter-gauge couplings,  the
action is quadratic in the gauge field components after imposing a
boundary condition. This is in contrast with the pure Chern-Simons
case where the action with the boundary condition imposed is linear
in one of the gauge fields and so allows one to trade the remaining
gauge fields in terms of the new degrees of freedom $U$ as in \eq{AU}.
In the ABJM case, one cannot see the emergence of new degrees of
freedom this way.
If one does try to integrate out some of the gauge
fields as in the original boundary condition
approach for the pure Chern-Simons action, one finds that the
remaining gauge field is dynamical and the action is nonlocal. 
The resulting theory is complicated and highly nontrivial. It is hard
to see how to proceed with this approach. 
On the other hand, one can
still follow the approach of adding degrees of freedom to derive the 
action for the boundary theory. In this case, one will still have
the added boundary gauge degrees of freedom $g$ and
the degrees of freedom $\l$ which arises from the gauge field $A$.
However, they do not
combine into a single gauge invariant variable anymore. 
The  $U(N)\times U(N)$ 
gauge symmetry is therefore manifest and will play an important role.
Another new feature is the presence of supersymmetry at the boundary.
This provides additional constraints on the form of the 
boundary degrees of freedom and their action.

\section{Action for Multiple Self-Dual Strings on M5-Branes}


\subsection{Bosonic ABJM open membrane theory with boundary} 

Let us apply our above construction to study the ABJM theory in the
presence of a boundary.  
The bosonic part of the action for the ABJM theory is given by
\be
S_{ABJM} = S_{CS} + S_C
\ee
where
\be
S_{CS} = \frac{k}{4 \pi} \int_M \Tr 
\left( \Aa d \Aa + \frac{2}{3}\Aa^3 \right)  
- \frac{k}{4 \pi} \int_M \Tr \left( \Ab d \Ab + \frac{2}{3} \Ab^3
\right), 
\ee
\bea
S_C =-\int_M \Tr \big(
D_M C_I^\dag D^M C^I \big) 
&-&  \frac{4\pi^2}{3 k^2} \int_M \Tr 
 \bigg(  C^I C_I^\dag C^J C_J^\dag C^K C_K^\dag 
+  C_I^\dag C^I C_J^\dag C^J C_K^\dag C^K \;\; \nn \\
 &+& 4  C^I C_J^\dag C^K C_I^\dag C^J C_K^\dag
- 6 C^I C_J^\dag C^J C_I^\dag C^K C_K^\dag 
\bigg)
\eea
and the level $k$ is a positive integer.
Here $A^{(1)}$ and $A^{(2)}$ are the gauge potentials for the two $U(N)$
factors in the gauge group. The matter fields $C^I$ ($I = 1,2,3,4$) are
in the bifundamental representation $(N, \bar{N})$ and the covariant
derivative acts as
$D_M C^I = \partial_M C^I + A^{(1)}_M C^I - C^I A^{(2)}_M$, where
$M=0,1,2$.

In the absence of a boundary, the action is gauge invariant under:
\be
A^{(i)} \rightarrow A^{(i)h^{(i)}} \;\; , \;\; 
C^I \rightarrow (h^{(1)})^{-1} C^I h^{(2)}.
\ee
But when there is a boundary, the Chern-Simons terms are not gauge invariant.
Naively, one may try to fix this by 
imposing boundary conditions as in the previous section. However there is 
an important difference here since the gauge fields are coupled
to the matter fields. As we discussed above, 
unlike the case for pure Chern-Simons action,
the bulk action is not linear in any component of the gauge fields.
Therefore, even after imposing boundary conditions, we will not be able to
integrate out a Lagrange multiplier.
If
we did follow this method, we would end up with a more complicated
action, and we would also not expect such an action to be equivalent at the
quantum level. Therefore, we would be left with the original action subject to
boundary conditions. 
The emergence of the needed degrees of freedom is obscure.
Instead, the second approach of introducing new degrees of freedom 
allows us to explicitly preserve
the desired symmetries of the bulk action, and interpret
the final invariant action in terms of boundary degrees of freedom.

To maintain gauge invariance, we can add boundary terms of the form described
in the previous section for each of the gauge fields $A^{(1)}$ and $A^{(2)}$.
The combined action would then be gauge invariant. Since the ABJM theory
also has conformal invariance, we should include the boundary kinetic terms
also. As a result,  the required boundary actions are
\be
\label{GWZW1}
S_{Bdry,1} = -\frac{k}{8\pi} \int_{\partial M} 
\Tr ( g^{-1}D^{(1)}_{\mu}g )^2 + \frac{k}{4\pi} \int_M 
(\omega_3(A^{(1)g}) - \o_2(A^{(1)}) )
\ee
and
\be
\label{GWZW2}
S_{Bdry,2} = -\frac{k}{8\pi} \int_{\partial M} 
\Tr ( \hat{g}^{-1}D^{(2)}_{\mu}\hat{g} )^2 
- \frac{k}{4\pi} \int_M 
(\omega_3(A^{(2)\hat{g}}) - \o_2(A^{(2)}) ).
\ee
Note that although the bulk Chern-Simons terms for $A^{(1)}$ and
$A^{(2)}$ 
differ by a
relative sign, as previously mentioned the sign of the boundary kinetic terms
in the actions (\ref{GWZW1}) and
(\ref{GWZW2}) is not fixed by conformal invariance. We have therefore chosen the
same sign for these terms so that we have conventional kinetic terms, and 
hence a 
well defined quantum field theory, for the
boundary fields $g$ and $\hat{g}$. More explicitly, we have
\be \label{SB1}
S_{Bdry,1} [g,A^{(1)}]
= S_{WZW}^{(-)}[g] + \frac{k}{4 \pi} \int_{\del M} \del_+ g g^{-1} A^{(1)}_- 
- \frac{k}{8 \pi} \int_{\del M} A^{(1)}_\m\,{}^2,
\ee
\be \label{SB2}
S_{Bdry,2} [\gh,A^{(2)}]
= S_{WZW}^{(+)}[\gh] - \frac{k}{4 \pi} \int_{\del M} \del_- \gh \gh^{-1} A^{(2)}_+ 
- \frac{k}{8 \pi} \int_{\del M} A^{(2)}_\m\,{}^2 ,
\ee
where the WZW actions are defined by
\be
S^{(\pm)}_{WZW}[h] := -\frac{k}{8\pi} \int_{\partial M} 
\Tr ( h^{-1}\partial_{\mu}h )^2 
\pm \frac{k}{12\pi} \int_M \Tr \left( h^{-1}dh \right)^3.
\ee

Adding the boundary actions \eq{GWZW1} and \eq{GWZW2} to the action for the
boundary fields $g$ and $\gh$, the full gauge invariant and 
conformal invariant action describing a system of $N$ ABJM 
open membranes is given by
\bea
S_T &=&  \frac{k}{4 \pi} \int_M \Tr 
\left( \Aa d \Aa + \frac{2}{3}\Aa^3 \right)  
- \frac{k}{4 \pi} \int_M \Tr \left( \Ab d \Ab + \frac{2}{3} \Ab^3
\right) \nn\\ 
&& -\int_M \Tr \big(D_M C_I^\dag D^M C^I \big) 
+  \frac{4\pi^2}{3 k^2} \int_M \Tr 
 \bigg(  C^I C_I^\dag C^J C_J^\dag C^K C_K^\dag 
+  C_I^\dag C^I C_J^\dag C^J C_K^\dag C^K \;\; \nn \\
&& \qquad \qquad \qquad \qquad \qquad \qquad 
+ 4  C^I C_J^\dag C^K C_I^\dag C^J C_K^\dag
- 6 C^I C_J^\dag C^J C_I^\dag C^K C_K^\dag 
\bigg) \nn\\
&& + S_{WZW}^{(-)}[g] + \frac{k}{4 \pi} \int_{\del M} \del_+ g g^{-1} A^{(1)}_- 
- \frac{k}{8 \pi} \int_{\del M} A^{(1)}_\m\,{}^2 \nn\\
&&+ S_{WZW}^{(+)}[\gh] - \frac{k}{4 \pi} \int_{\del M} 
\del_- \gh \gh^{-1} A^{(2)}_+ 
- \frac{k}{8 \pi} \int_{\del M} A^{(2)}_\m\,{}^2 \; -\int_{\del M}  V_{\del M}(C).
\label{S-T}
\eea
Here we have included a possible boundary interaction term $V_{\del M}$
for the matter fields $C^I$. Due to conformal invariance, 
the boundary potential is 
quartic, $V_{\del M}
\sim \sum C^4$. Classically, it is
\be \label{VBdry}
V_{\del M} = \a^{IJKL} \Tr (C^I C_J^\dag C^K C_L^\dag). 
\ee
The coefficients  satisfy
$\a^{IJKL}{}^\dag = \a^{JILK} = \a^{LKJI}$.
Classically these coefficients are free. 
The requirement of quantum conformal invariance 
will provide further constraints on these coefficients. The form of $V_{\del M}$
is further constrained  
if the open membranes theory has a boundary which preserves some amount of
supersymmetry.

We can now see how this formulation encodes boundary conditions for the gauge
fields, in the form of boundary equations of motion. 
To be more explicit we consider the boundary to be at
$x^2 = 0$. It is important to note that variations of
the gauge fields in $S_C$ will contribute to the bulk equations of motion, but
they will not give rise to any boundary terms. Therefore
using the results from section~\ref{CSBDoF}, we obtain immediately 
the boundary condition from the variation $\d A^{(1)}$:
\be
A^{(1)g}_{+} = 0
\ee
which is equivalent to
\be \label{Ap}
A^{(1)}_{+} = - (\partial_{+}g) g^{-1}.
\ee
Clearly a similar analysis would follow for $A^{(2)}$, but the 
different relative
sign between the bulk and boundary terms would give the result
\be
A^{(2)\hat{g}}_{-} = 0,
\ee
or equivalently
\be \label{Am}
A^{(2)}_{-} = - (\partial_{-}\hat{g}) \hat{g}^{-1}.
\ee

Note that in the bulk action, compared to the Chern-Simons action there are
variations of $A^{(i)}$, arising from the gauged kinetic terms for $C$. 
These do not
modify the boundary terms, but change the bulk equations of motion, so that
instead of $F^{(i)}=0$ we have the  equation of motion
\bea \label{F-bulk}
\frac{k}{2\pi} F^{(1)} & = & * C^IDC^{\dagger}_I - h.c. \nn \\
\frac{k}{2\pi} F^{(2)} & = & *  DC^{\dagger}_IC^I - h.c.
\eea
This is a direct consequence of the fact that, due to the presence of matter, 
the ABJM theory is not topological.

Next we consider the variations of $g$ and $\hat{g}$. Since these boundary
fields only appear in the modified Chern-Simons actions, their effect is
described fully by the considerations of section~\ref{CSBDoF}. The boundary
equations of motion therefore result in the boundary conditions
\be \label{Fmn}
F_{\m\n}^{(i)} = 0, \qquad \m,\n =0,1. 
\ee 
Together with \eq{Ap} and \eq{Am}, this means
\be
A^{(1)}_\m = - (\partial_{\m}g) g^{-1} \quad 
\mbox{and}\quad
A^{(2)}_\m = - (\partial_{\m}\gh) \gh^{-1}.
\ee

Finally, variations of $C^I$ give the boundary condition
\be \label{bc-C}
D_2 C^I =  - U^I,
\ee
where $U^I$ denotes the variation 
$ \d V_{\del M}/ \d C_I^\dag$.  Using this and the
consistency of \eq{Fmn} with \eq{F-bulk} implies that
\be \label{UC}
C^I U_I{}^\dag - U^I C_I{}^\dag =0
\ee 
on $\del M$. It is easy to check that \eq{UC} is satisfied for
\eq{VBdry} and so the bulk equations of motion \eq{F-bulk} are consistent
with the boundary conditions \eq{Fmn}.

\subsection{$\cN = (4,4)$ boundary ABJM theory}

In the previous section we did not include supersymmetry in the discussion.
If we consider a set of multiple M2-branes ending orthogonally on a M5-brane
in flat space, the intersecting M2/M5 system will preserve a quarter of the
supersymmetry, i.e.\ $\cN = (4,4)$ in two dimensions. For $k>2$ the
$C^4/Z_k$ orbifold
will break the supersymmetry further, but we expect the boundary field content
to be consistent with the $\cN = (4,4)$ multiplet structure for 
any value of $k$, in the same way as the ABJM field content is 
consistent with $\cN=8$ supersymmetry in three
dimensions. Since the bulk ABJM theory has only manifest $\cN=6$ supersymmetry,
we expect the boundary action, upon imposing a suitable boundary condition on
the $C$-fields which corresponds to the M5-brane, to have only manifest 
$\cN = (3,3)$ supersymmetry. 

First, let us look at the boundary condition for the $C$-fields. 
Following the 
$\cN=(2,2)$ superspace construction of \cite{Kl}, 
it is convenient to introduce
$SU(2)$ fields by writing $C^I=\{ Z^I, W^I{}^\dag \}$, $C_I^\dag=\{
Z_I^\dag, W_I \}$, where the index $I$ on the right hand side now runs
from 1 to 2. Here $Z$'s are in the representation $(N, \bar{N})$ and
$W$'s are in the representation $(\bar{N}, N)$. In this formulation, the
ABJM theory posesses 
an  $SU(4)_R$  global symmetry. 
A particular BPS
configuration (`D-term' type) of the ABJM theory
has been considered in \cite{lin}:
\bea 
\frac{k}{2\pi} D_2 Z^I +  Z^I (Z^\dag Z - W W^\dag) - (Z Z^\dag - W^\dag
W) Z^I =0,  \label{bc-ZWa}\\
- \frac{k}{2\pi} D_2 W^I{}^\dag +  W^I{}^\dag (Z^\dag Z - W W^\dag) 
- (Z Z^\dag - W^\dag W) W^I{}^\dag =0,  \label{bc-ZWb}
\eea
and
\be \label{bc-ZW-x1}
F_I^\dag:= \frac{4\pi}{k} \e_{IJ} \e^{KL} W_K Z^J W_L =0, \quad 
G^\dag{}^I := \frac{4\pi}{k} \e^{IJ} \e_{KL} Z^K W_J Z^L = 0. 
\ee
If one identifies these equations with the boundary condition \eq{bc-C},
one can deduce further details of the boundary potential $V_{\del M}$.
We obtain 
\be
V_{\del M, D} =  \frac{\pi}{k} \Tr [ 
(Z Z^\dag - W^\dag W)^2 - (Z^\dag Z - W W^\dag)^2 ] ,
\ee
where the full potential can include an additional term which 
vanishes when \eq{bc-ZW-x1} are imposed. 
Another BPS configuration ('F-term type') 
\cite{lin}
is given by
\bea
\frac{k}{4\pi} D_2 Z^I - \e^{IJ} \e_{KL} W^\dag{}^K Z^\dag_J W^\dag{}^L =0, 
\label{bc-ZWc}\\
\frac{k}{4\pi} D_2 W^\dag{}^I - \e^{IJ} \e_{KL} Z^K W_J Z^L =0,
\label{bc-ZWd}
\eea
and
\be \label{bc-ZW-x2}
N^I:=\s C^I - C^I \hat{\s} =0, \quad I=1,\cdots, 4,
\ee
where
$\s:= 2\pi/k (Z Z^\dag -W^\dag W)$, $\hat{\s}:= 2\pi/k(Z^\dag Z - W
W^\dag)$.
This  corresponds to the  boundary potential
\be
V_{\del M, F } = - \frac{2 \pi}{k} \Tr [\e_{IJ} \e^{KL} Z^I W_K Z^J W_L] +
h.c. , 
\ee
where the full potential can include an additional term which 
vanishes when \eq{bc-ZW-x2} is imposed. 
These boundary potentials have also been considered in \cite{Ber2}.
Here we will argue that the full boundary potential is given by
\be \label{V-full}
V_{\del M} := V_{\del M, D} + V_{\del M, F}.
\ee
To see this, it is sufficient that in principle one may add to
\eq{V-full} a term  which vanishes when either \eq{bc-ZW-x1} or
\eq{bc-ZW-x2} are imposed. However, combined with conformal invariance,
there is no such quartic polynomial one can construct. 

Before we move on, we remark that
we have adopted above the definitions of \cite{Kl} for $F_I, G^I,
N^I, \s, \hat{\s}$. Using this notation, the potential for $C^I$
in the ABJM theory can be written as 
\be
V_M = 
\Tr [F_I^\dag F^I + G^\dag{}^I G_I] 
+ \Tr[N_I^\dag N^I ]. 
\ee

Next, let us consider the boundary actions \eq{SB1} and \eq{SB2} 
we have added to the bulk ABJM action. Obviously these cannot be the 
whole story as the fields $g$, $\gh$ we added do not 
fully describe the bosonic content of a multiplet of
$\cN=(4,4)$ supersymmetry, which consists of 4 real scalar 
and 4 Weyl fermionic degrees of freedom. Therefore we need to
supersymmetrize our boundary action by supplementing it with
additional fields.  
We will show now how to do this. 

The supersymmetric WZW action is a particular type of 
nonlinear sigma
model in 2-dimensions. This type of nonlinear sigma 
model generalizes the original supersymmetric
construction of nonlinear sigma models \cite{zumino,AGZ} by utilizing
in addition to a metric also a 2-form. The action is entirely
determined by a flat connection whose torsion is determined by the
2-form \cite{SSTV1,SSTV2,SSTV3}.  The most general such manifolds are given by
semisimple Lie groups. With respect to a basis of left invariant
one-forms, the metric is simply given by the constant $\d_{ab}$ and
the torsion is given by the group structure constants $f_{abc}$.
The relative size between these two terms is fixed by supersymmetry. This
gives rise to supersymmetric WZW models. These
models always have $\cN= (n,n)$ supersymmetry and $\cN=(4,4)$
supersymmetry is the highest one can get. Moreover  $\cN=(3,3)$
implies $\cN=(4,4)$. Therefore although we may only expect $\cN=(3,3)$ for the
full boundary ABJM theory, this WZW 
sector of the theory will actually have $\cN=(4,4)$
supersymmetry.
Futher conditions of supersymmetry impose additional contraints on the form
of the group manifold. For  $\cN = (4,4)$, 
these are a particular type of quaternionic group
manifolds. A list of all possible such
group manifolds is given in  table 1 of
\cite{SSTV1}, or by taking products of
factors there. 

The $\cN=(4,4)$ WZW theory we are after has $SU(2)\times SU(2)$
R-symmetry. It is instructive to recall how the R-symmetry is realized
in the $\cN=(4,4)$ models in general. 
Let us first consider the case of $\cN=(3,3)$ WZW model which has
a $SU(2)$ R-symmetry. 
The simplest such WZW model is given by \cite{IV1}
\be
S = \int \del_+ u \del_- u + \cL_{WZW}(q_\a{}^\b) + i \xi_+^{\a a}
\del_- \xi_{+\a a} 
+ i \xi_-^{\a a} \del_+ \xi_{-\a a},
\ee
where $u, q_\a{}^\b$ are bosonic and $\xi_\pm$ are fermionic $U(1)$
fields and  $\a, \b= 1,2$,  $a,b=1,2$. 
The
indices $\a, \b$ are acted on by the R-symmetry $SU(2)_1$. Since we
know $\cN=(3,3)$ implies $\cN=(4,4)$, there must be a second $SU(2)_2$
R-symmetry. Indeed, this is given by 
the $SU(2)_2$ which acts on the  indicies $a,b$ of the fermions. We note
that this second $SU(2)_2$ cannot be seen if one considers only the bosonic
sector. We also note that this pattern of R-symmetry enhancement is
generic and applies to the more general situation where the fields are
nonabelian. Now back to the $\cN=(4,4)$ case. 
Since in general the group manifold $G$ should contain
the R-symmetry as a subgroup,
in order to have an R-symmetry $SU(2)_1\times SU(2)_2$
which is fully visible in the bosonic sector, one has to consider a
product group manifold $G= G_1 \times G_2$ with $SU(2)_1$  acting on
the factor $G_1$ and the $SU(2)_2$ acting on $G_2$. This product
structure of the group manifold fits with our boundary degrees of
freedom $g\in U(N)_1 , \gh \in U(N)_2$ 
introduced above. Inspecting table 1
of \cite{SSTV1} suggests that six additional $U(N)$ degrees of
freedom should be introduced in such a way that three of the new fields
combine with $g$ to form a group $G_1= U(2N)_1$ and the other three new 
fields combine with $\gh$ to form a group $G_2 = U(2N)_2$.
That is, the required $\cN=(4,4)$ supersymmetric WZW model should be based
on the group manifold
\be \label{GG}
G_1 \times G_2 = U(2N) \times U(2N).
\ee
This is the minimal group manifold with the property that the
commutant of the R-symmetry is $U(N)\times U(N)$. Explicitly, one can
denote the  group elements $g \in G_1$, $\gh \in G_2$ as
\be \label{gghat}
g = \exp( u \id + \varphi_i \s^i), \quad
\gh = \exp( \uh \id + \hat{\varphi}_i \s^i), 
\ee  
where $\id$ is the identity $2 \times 2$ matrix and $\s^i$ are the
Pauli matrices. In this representation, the $SU(2)_1$ R-symmetry, for
example,  is
represented by the generators
\be
\id_{N\times N} \otimes \s_i
\ee
in $U(2N)_1$.
The commutant subgroup is the one which 
one can gauge by coupling to the bulk ABJM gauge fields. Therefore it 
must contain $U(N) \times U(N)$. The choice \eq{GG} is a minimal
choice because the commutant of the R-symmetry is exactly $U(N) \times
U(N)$.  We will comment on
the possibility of other nonminimal choices of group manifold later. 

Having successfully supersymmetrized the boundary WZW action to
have the desired $\cN=(4,4)$ supersymmetry, we now come to the issue
of the cancellation of the gauge
noninvariance of the boundary Chern-Simons action.
It is easy to see that our previous construction can be carried out
in exactly the same way by simply embedding $U(N)\times U(N)$ into
$U(2N) \times U(2N)$, i.e. we tensor all the ABJM fields with
$\id_{2\times 2} \times \id_{2\times 2}$. With this interpretation,
the bosonic action of the $\cN=(4,4)$ boundary ABJM theory with $U(N)
\times U(N)$ gauge group is given by \eq{S-T},
with the definition of $\Tr$ including an additional normalization
factor of 1/2. This obviously leaves the bulk theory unchanged. As for
the boundary theory, the normalization of the WZW term is constrained
by a topological argument \cite{W2}. Note that this constrains
only the $u$ and $\hat{u}$-part (as defined in \eq{gghat}) 
and we do get the right
normalization. 

Some comments on this supersymmetrized action follow.

\bit
\item[1.] One can
certainly consider other $\cN=(4,4)$ group manifolds in which \eq{GG} is
embedded,
for example $G_1 \times G_2 =  U(2M) \times U(2M)$ with large enough
$M$ \footnote{
All the groups in the table 1 of \cite{SSTV1} can be embedded in
$U(2M)$ for sufficiently large $M$ except for the exceptional groups.
However these can be included only as direct products which will decouple
from the other fields.  
}. In this case, the bulk plus boundary action can be
obtained as a subsector of our construction with a bulk 
 $U(M) \times U(M)$ ABJM theory. 

\item[2.] In the above, we have  supersymmetrized the WZW actions of
the boundary actions \eq{SB1}, \eq{SB2} with $\cN=(4,4)$
supersymmetry. To render the whole boundary action supersymmetric,
one still needs to supersymmetrize the $gA$ and $AA$ type of terms with
the accompanying fermion action. 
We expect the resulting action will be superconformally invariant at 
the quantum level.

\item[3.] The coefficient of the kinetic term of $g$ in
the boundary action \eq{SB1} was originally fixed by requiring a WZW
action is formed so that one has conformal invariance. Here we see
that the coefficient is also fixed by requiring supersymmetry of
the nonlinear sigma model.
\eit

\subsection{Multiple self-dual strings action}

\if 
To start with, it is
convenient to first rewrite the action 
\eq{S-T} for $S_T$ by using  rescaled fields. 
Let the membrane tension be
$T_3$, where $T_3$ is of the order of $1/l_P^3$. Introducing the rescaled
variables $C= \sqrt{T_3} \cC$, 
the $C$-dependent part of the action \eq{S-T} for $S_T$ is
\be \label{SC}
-T_3 \int_M (D \cC)^2 - T_3^3 \int_M V_{ M}(\cC) 
-T_3^2 \int_{\del M} V_{\del M}(\cC).
\ee 
\fi
 
To derive the action for the multiple self-dual strings, 
let us
consider a bunch of open M2-branes suspended between two parallel
M5-branes separated by a distance $x^2=L$. 
Denoting the string worldsheet coordinates 
by $(\s_0, \s_1)$,  one has to take a limit such that
all the physical configurations on the M2-branes are independent of
$\s_2$. 
This corresponds to suppression of  the
membranes modes of motion.
This can be achieved by taking all the gauge fields to be independent of
$\s_2$ and the bulk matter fields 
to be independent of $\s_2$ covariantly:
\be
\del_2 A_M =0, \quad D_2 C^I =0, 
\ee 

The fact that the M2-branes are ending on an
M5-brane means that the four fields $C^I$ should be divided into two
groups, with two $C$'s describing the longitudinal directions on the
M5-brane, and two $C$'s describing the directions transverse to the
M5-brane. Denoting the latter as $W^I \propto \id$ 
and the former as $Z^I$, $I=1,2$,
the $C$-dependent part of the action \eq{S-T} becomes
\be \label{VV}
-L \int d^2 x  \bigg[ |D Z|^2 +  V_{ M}(Z) + \frac{1}{L}
V_{\del M} (Z) \bigg].
\ee
Here we have taken a gauge $A_2^{(i)} =0$ so that $C^I$ are independent
of $\s_2$ and \eq{VV} makes sense.

For the Chern-Simons terms, it is easy to see that there is nothing left after
dropping derivatives with respect to $\s_2$ and setting $A_2^{(i)}=0$.
As for the WZW terms, since the M2-branes end on two different
M5-brane, we get an $\cN=(4,4)$ supersymmetric $U(2N)\times U(2N)$ 
WZW action on each boundary. Consider one of the boundaries, say $\del
M_1$. Adding the boundary action to  \eq{VV}, we obtain the  
action for $N$ multiple self-dual strings living on the boundary
$\del M_1$,  
\bea \label{S-final}
 && -L 
\int \big[ |D Z|^2 + V_{ M}(Z) + \frac{1}{L}
V_{\del M} (Z) \big]\nn \\
&& + 
S_{WZW}^{(-)}[g] + \frac{k}{4 \pi} \int  \del_+ g g^{-1} A^{(1)}_- 
- \frac{k}{8 \pi} \int  A^{(1)}_\m\,{}^2 \\
&&+  
S_{WZW}^{(+)}[\gh] - \frac{k}{4 \pi} \int  \del_- 
\gh \gh^{-1} A^{(2)}_+ 
- \frac{k}{8 \pi} \int  A^{(2)}_\m\,{}^2 
+ \mbox{fermions}.\nn
\eea
We note that since $A^{(1)}_\pm$ and  $A^{(2)}_\pm$ appear linearly in the action,
one may try to integrate out, for example, $A^{(1)}_-$ and $A^{(2)}_+$ and
obtain constraints between $A^{(1)}_+$, $A^{(2)}_-$ with $Z, g$ and
$\gh$. Solving these constraints for $A^{(1)}_+$ and $A^{(2)}_-$  
and substituting back into
\eq{S-final} would give us an action given in terms of $Z, g$ and
$\gh$ only.
The constraints are however complicated and solving it involves
nonlocal expressions. So it is better to present the action in the
form \eq{S-final}. We also note that the introduction of the scale $L$
breaks the conformal symmetry of \eq{S-T} and the 
action \eq{S-final} describes self-dual strings
configurations that are at energy scale below $1/L$. It is interesting
to ask what \eq{S-final} flows to at lower energies. Addressing this 
is beyond the scope of this paper.

\section{Discussion}

In this paper we have provided a new method to treat a twisted Chern-Simons
matter system with boundary. The guiding principle in our construction
is the manifest preservation of gauge symmetry and conformal symmetry
of the system including the boundary. By applying our method to the
ABJM theory with boundary, 
and together with the requirement of respecting the 
$\cN=(4,4)$ supersymmetry multiplet structure, 
we identified the new degrees of freedom $g$ and 
$\gh$
that must be present on the worldsheet of multiple self-dual strings. 
These degrees of freedom generate a 
$U(2N) \times U(2N)$
Kac-Moody current algebra  on
the worldsheet. It will be interesting to understanding better the
role of these currents in the physics of multiple self-dual strings. For
example, these currents could tell us something about intersecting brane
configurations on M5-branes.
It is also  an open question whether these currents couple to non-abelian gauge
bosons in the background.

Another result we obtain is the determination of the
boundary potential $V_{\del M}$ with the use of supersymmetry and
the scaling property of the potential. It would be interesting to consider
the BPS equations for \eq{S-final} and study the properties of other
solitons within M5-branes.

In this paper, we have considered a system of free self-dual strings. It
will be interesting to include couplings to background fields on the
M5-brane. One particularly interesting background is the non-abelian
self dual 2-form potential which is expected to arise when there are multiple
M5-branes. The description of a non-abelian tensor is an open problem. An
interesting attempt has been considered in \cite{Gu} which involves the
use of a loop space. The origin of this ``extra dimension'' is however
not clear. One can speculate that this is 
related to the difficulty of quantizing the membrane which has 
a continuous spectrum. Consequently this may lead to new issues in
reducing the membrane action to the string action.

In \cite{Ber2}, chiral WZW actions with opposite signs of kinetic terms 
were obtained on the boundary of the open M2-branes theory. 
It was argued that a parity operation can be defined which exchanges the
two kinetic terms of the WZW actions, resulting in a nonchiral theory. 
However this parity operation does not address the issue of the
ill-defined kinetic terms.
In our construction, we obtain nonchiral WZW actions immediately.
Moreover our WZW action has well defined kinetic terms. 

By dimensional reduction, one can obtain the action for multiple self-dual 
strings on the NS5-brane. This action would seem to be different from what 
one would obtain by considering D2-branes suspended between NS5-branes.
In particular, the D2-branes non-abelian Born-Infeld action does not
contain any Chern-Simons terms and so it would not appear necessary to
introduce the boundary degrees of freedom $g$ and $\gh$.
However we predict that the WZW action will still arise in the
boundary theory. 
 It would  
be interesting to understand this issue properly, as well as the
reduction to the D4-brane system.

Finally we comment that one 
may also consider having a set of $N$
M2-branes ending on two separated M9-branes as in the Horava-Witten
setup \cite{HW}.  We expect the $E_8\times E_8$ gauge symmetry will
arise from the gravitational anomaly in the same way as in \cite{HW}.
The  Kac-Moody symmetry on the worldsheet is however new and  
seems to suggest the
emergence of additional gauge symmetry in spacetime. This is
entirely unclear. It is important to understand better the role of the
Kac-Moody symmetry, both for the multiple self-dual strings on M5-brane
and for the multiple heterotic strings, which is one of the main predictions of our
construction. 

\section*{Acknowledgements}

It is a pleasure to thank Nick Dorey, Pei-Ming Ho, Chris Hull, Valya V. Khoze 
and Yutaka Matsuo
for discussions. We also thanks the organizers of the Corfu
Summer Institutes 2009  for hospitality where this work was finalized.
CSC acknowledges EPSRC and STFC for support. 
DJS acknowledges STFC for support.


\end{document}